\documentclass[letterpaper, 10 pt, conference]{ieeeconf}  
\IEEEoverridecommandlockouts
\usepackage{algorithm}
\usepackage{fnpct}
\usepackage{array}
\usepackage{tabularx}
\usepackage{footnote}
\usepackage{booktabs}
\usepackage{multirow}
\usepackage{makecell}
\usepackage{csquotes}
\usepackage{pgfplots}
\usepackage{subcaption} 
\usepackage{caption}
\usepackage{xcolor}
\usepackage{graphics} 
\usepackage{graphicx} 
\usepackage{epsfig} 
\usepackage{times} 
\usepackage{amsmath} 
\usepackage{amssymb}  
\usepackage{algorithm}
\usepackage{url,verbatim,color,comment}
\usepackage[noend]{algpseudocode}
\usepackage{eucal}
\usepackage{amsfonts}
\usepackage{fancyhdr}
\usepackage{mathtools}
\usepackage{placeins}

\usepackage{flushend}
\usepackage{tikz}
\pgfplotsset{compat=newest}

\makeatletter
\let\NAT@parse\undefined
\makeatother
\usepackage[style=ieee,natbib=true, url=false, doi=false,  mincitenames=1, maxcitenames=1]{biblatex}
\usepackage[hidelinks]{hyperref}
\hypersetup{colorlinks,citecolor= violet,linkcolor=blue, urlcolor=black}

\title{\LARGE \bf
Optimizing Falsification for Learning-Based Control Systems:\protect\\ A Multi-Fidelity Bayesian Approach
}

\author{Zahra Shahrooei$^{1}$, Mykel J. Kochenderfer$^2$, and Ali Baheri$^1$
\thanks{$^{1}$Zahra Shahrooei and Ali Baheri are with the Department of Mechanical Engineering at Rochester Institute of Technology.
        {\tt\small zs9580@rit.edu} and {\tt\small akbeme@rit.edu}}%
\thanks{$^{2}$Mykel J. Kochenderfer is with the Department of Aeronautics \& Astronautics at Stanford University. 
        {\tt\small mykel@stanford.edu}}
}

\begin{document}

\maketitle

\thispagestyle{empty}
\pagestyle{empty}

\begin{abstract}
Testing controllers in safety-critical systems is vital for ensuring their safety and preventing failures. In this paper, we address the falsification problem within learning-based closed-loop control systems through simulation. This problem involves the identification of counterexamples that violate system safety requirements and can be formulated as an optimization task based on these requirements. Using full-fidelity simulator data in this optimization problem can be computationally expensive. To improve efficiency, we propose a multi-fidelity Bayesian optimization falsification framework that harnesses simulators with varying levels of accuracy. Our proposed framework can transition between different simulators and establish meaningful relationships between them. Through multi-fidelity Bayesian optimization, we determine both the optimal system input likely to be a counterexample and the appropriate fidelity level for assessment. We evaluated our approach across various Gym environments, each featuring different levels of fidelity. Our experiments demonstrate that multi-fidelity Bayesian optimization is more computationally efficient than full-fidelity Bayesian optimization and other baseline methods in detecting counterexamples. A Python implementation of the algorithm is available at \url{https://github.com/SAILRIT/MFBO_Falsification}.

\end{abstract}

\section{Introduction and Related Works}

The rapid development of cyber-physical systems, including autonomous vehicles and robotics, has enhanced the need for controllers to function both safely and reliably. Ensuring the safety of these systems is crucial to prevent catastrophic failures. This involves using simulation-based tests to identify potential failures, determine the most likely failure mode, and estimate system failure likelihood. Various methods have been developed for generating test scenarios for autonomous vehicles, including knowledge-based \cite{da2024ontology,li2020ontology}, data-driven \cite{o2018scalable,esenturk2021analyzing}, and search-based approaches \cite{feng2020testing,klischat2020synthesizing}. Knowledge-based methods use formal models to create scenarios based on traffic rules, while data-driven approaches rely on real-world driving data to identify critical situations. On the other hand, search-based techniques explore possible scenarios by systematically testing different driving conditions to ensure comprehensive coverage.

Falsification is a search-based testing procedure that tries to disprove safety by discovering an input (a counterexample) that leads to a safety specification violation. A safety specification is a set of requirements that must be met by the system during execution. For falsification, the specification is often expressed in signal temporal logic \cite{donze2010robust} or metric interval temporal logic \cite{fainekos2009robustness}. Various algorithms have been developed to solve the falsification problem. Among these, optimization-based algorithms are designed to guide the falsification process by defining a suitable objective function that guides the search towards an environment trace where the specification is not satisfied. For tackling this global optimization problem, black-box methods offer a structured approach for uncovering safety failures in systems without the need for understanding their internals. Black-box testing involves searching the uncertainty space, where the parameters are system inputs and the goal is to detect failures. Sequential search algorithms include ant colony optimization \cite{annapureddy2010ant}, stochastic local search \cite{deshmukh2015stochastic}, and simulated annealing \cite{aerts2018temporal}. While these algorithms excel at efficiently navigating the uncertainty space, they do not make effective use of the information gained from prior simulations.

In this paper, we use Bayesian optimization (BO), a black-box algorithm to globally optimize an unknown objective function by approximating its model using Gaussian processes (GPs) \cite{williams2006gaussian} and continuously updating posterior estimates to help determine the next point to evaluate. BO has seen successful applications ranging from robotics \cite{berkenkamp2016safe, jacobs2022framework} and control \cite{baheri2017real} to design optimization problems \cite{garnett2010bayesian, baheri2018iterative} and hyper-parameter tuning in machine learning \cite{frazier2018tutorial}. Additionally, BO has been extensively applied in safety testing \cite{deshmukh2017testing, gangopadhyay2019identification, ghosh2018verifying, mathesen2021efficient, ramezani2022falsification, 10620179}. These studies typically focus on safely optimizing an unknown function \cite{kim2020safe}. For instance, \citet{ghosh2018verifying} use BO to address the falsification problem in uncertain closed-loop control systems. They model the unknown specification by breaking it down into a parse tree with individual constraints represented by nodes and each constraint modeled using GPs to reduce number of the required queries for falsification. \citet{deshmukh2017testing} proposed the use of random embedding BO (REMBO) \cite{wang2016bayesian} to tackle the falsification problem in cyber-physical systems with high-dimensional input spaces. \citet{ramezani2022falsification} applied two other extensions of BO to falsification, trust region BO (TuRBO) \cite{eriksson2019scalable} and $\pi$BO \cite{hvarfner2022pi}. TuRBO makes use of several local GPs, each within its trust region that shrinks or expands based on the success or failure of solutions. $\pi$BO, on the other hand, focuses on injecting the expert prior knowledge of the falsifying points to the BO framework. All these studies apply BO directly on the full-fidelity simulator. A significant drawback of using BO for simulation-based falsification is the high computational cost of evaluating the objective function on full-fidelity simulators.

Multi-fidelity BO is a promising approach for cost reduction when leveraging rich, high-fidelity simulators. This framework has been successfully applied to engineering design problems, scientific discovery, and hyperparameter optimization \cite{tran2020multi, wu2020practical, charayron2023multi}. Multi-fidelity BO exploits information from various fidelities to achieve higher efficiency than using full-fidelity. Fidelity level in simulators refers to how closely a simulator replicates the real system. High-fidelity and low-fidelity simulators can differ widely based on various factors across different applications. For instance, model fidelity refers to the simplification of the mathematical representation of the physical system, typically by simplifying the differential equations being solved or the numerical model. Additionally, sensor fidelity is related to having more accurate sensors within the simulator while environmental fidelity is associated with considering the effect of environment factors such as weather condition or disturbances in simulator.


There are also works on multi-fidelity BO in control systems \cite{beard2022safety, baheri2023safety, marco2017virtual, baheri2023exploring}. For example, \citet{marco2017virtual} incorporate two degrees of fidelity involving both simulations and physical experiments in order to optimize controller parameters more efficiently. They propose an acquisition function aimed at striking a balance between cost and accuracy. However, they assume that the costs of simulation and real-world experimentation are predefined constants. Hence, the proposed algorithm suffers from lack of robustness to cost variation.

This paper is an extension of our conference publication \cite{shahrooei2023falsification} that proposes the first multi-fidelity BO application to the computationally demanding task of falsification for closed-loop control systems using data from different levels of fidelity. The framework has been shown to successfully cut down the cost of falsification in comparison to using full-fidelity simulator. However, we used predefined costs for the simulators with varying levels of fidelity, chosen manually and without any systematic measurement or analysis even though these costs directly influence the acquisition function and, consequently, the selection of the optimal fidelity level for evaluation. This paper expands the results of \cite{shahrooei2023falsification} through the
following contributions:
\begin{itemize}[]
\item Unlike the previous paper, which focused solely on a two-fidelity setting, we investigate both two-fidelity and three-fidelity levels. We explore the role of a third fidelity level in reducing the cost of falsification, enhancing the applicability and flexibility of our approach. Additionally, we provide a detailed analysis of the cost changes associated with transitioning from a two-fidelity setting to a three-fidelity setup, highlighting the trade-offs involved.
\item To avoid using manually assigned costs associated with different fidelity levels which can degrade the multi-fidelity BO performance, we propose a method to calculate the cost proportions of various fidelity degrees by measuring specific parameters, which offers a more realistic and adaptable cost model making multi-fidelity BO framework more robust and reliable in various applications.

\item Unlike previous work, we extend the use of multi-fidelity BO to falsification of complex systems, specifically in the context of autonomous driving to demonstrate the applicability of multi-fidelity BO approach.



\item We study the trade-off between cost and accuracy in the context of falsification. We demonstrate that using multi-fidelity BO can improve accuracy compared to using low-fidelity simulation alone and assess the efficacy of multi-fidelity BO in cost reduction compared to standard BO and other extensions of BO that rely exclusively on high-fidelity simulators in several case studies.

\end{itemize}

This paper is organized as follows. Section \ref{sec:prelim} specifies the problem formulation in terms of safety requirements and outlines the BO framework. Section \ref{sec:MFBO} explains how we use multi-fidelity BO for falsification. Section \ref{sec:Evaluation} describes how we implement multi-fidelity BO on four case studies. Section \ref{sec:result} discusses our empirical results. Section \ref{sec: Conclusion} concludes our work and provides potential future directions.

\section{Problem Formulation and Preliminaries}
\label{sec:prelim}

This section briefly reviews our notation, system safety specification, quantitative semantics, and BO concepts.

\subsection{Problem Formulation}
We assume that we have access to a set of simulators with varying degrees of fidelity specified by $S_{1}$ to $S_{q}$ for a system of interest $S$. All these $q$ simulators operate in a stochastic environment $\mathcal{E}$ that includes various types of uncertainty. We can parameterize each simulator with a set of environment parameters $\mathbf{e} \in \mathcal{E}$. Hence, each simulator takes an instance of the environment $\mathbf{e}$ as an input and returns a finite-length trajectory $\xi\left ( \mathbf{e} \right )$. Also, we have a controller $\pi$ trained for the highest-fidelity simulator $S_{q}$. We aim to test this controller performance against the system specification $\varphi$.

For this purpose, we use quantitative semantics to estimate the robustness value of safety specification satisfaction (dissatisfaction) $\rho _{\varphi }\left ( \xi \left ( \mathbf{e} \right ) \right )$. To falsify the controller, we explore search space for cases that correspond to negative robustness values ($ \rho _{\varphi }\left ( \xi \left ( \mathbf{e} \right ) \right )< 0$). This can be considered as an optimization problem with the robustness value as the objective function
\begin{equation}
    \underset{\mathbf{e}}{\mathrm{argmin}}\, \rho _{\varphi }\left ( \xi \left ( \mathbf{e} \right ) \right ) \label{(1)}
\end{equation}

Typically, solving this optimization problem using high-fidelity data is costly, but we combine the information from simulators with different levels of fidelity with limited experiments on the highest-fidelity simulator to identify the environment parameters that are likely to be counterexamples. In particular, we use BO to effectively direct the search over the environment and choose which simulator to use.

\subsection{Safety Specification and Quantitative Semantics}
A specification $\varphi$ is defined with respect to system trajectories. We conceptualize $\varphi$ as the set of all finite-horizon trajectories of the system that meet the system-level safety specification. A system trajectory $\xi_{S}\left (\mathbf{e} \right )$ for an environment parameter $\mathbf{e}$ satisfies the specification if and only if $\xi_{S}\left ( \mathbf{e} \right )\in \varphi $. In fact, $\varphi$ evaluates a system trajectory to true if $\xi_{S}\models \varphi$. The specification is composed of several individual constraints known as predicates. These predicates serve as the fundamental building blocks of the logic, which can be joined together using specific logical operations. We assume that a predicate $\mu$ is a smooth function along a system trajectory $\xi_{S}\left ( \mathbf{e} \right)$. We say a predicate is satisfied if $\mu \left ( \xi_{S}\left ( \mathbf{e} \right) \right )$ is greater than $0$ or falsified otherwise. The logical operations $\neg$ (negation), $\wedge$ (conjunction, which means \enquote{and}), and $\vee$ (disjunction, which means \enquote{or}) are used to combine these predicates. Rather than just checking the Boolean satisfaction of a predicate, the concept of robust or quantitative semantics is defined to quantify the degree of satisfaction. This allows us to move beyond binary notions of \enquote{safe} or \enquote{unsafe} by associating each predicate with a real-valued function $\rho _{\mu }\left ( \xi_{S} \left ( \mathbf{e} \right ) \right )$ of the trajectory $\xi_{S}\left ( \mathbf{e} \right )$, which provides a \enquote{measure} of the margin by which $\mu\left ( \mathbf{e} \right )$ is met. This robustness value $\rho _{\mu }\left ( \xi_{S} \left ( \mathbf{e} \right ) \right )$ is defined such that
\begin{equation}
 \xi_{S}\left (\mathbf{e} \right )\models \mu \leftrightarrow \rho _{\mu }\left (  \xi_{S}\left ( \mathbf{e} \right  )\right )> 0 \label{(2)}
\end{equation}

\begin{figure}[ht!]
     \centering
     \includegraphics[scale=0.4]{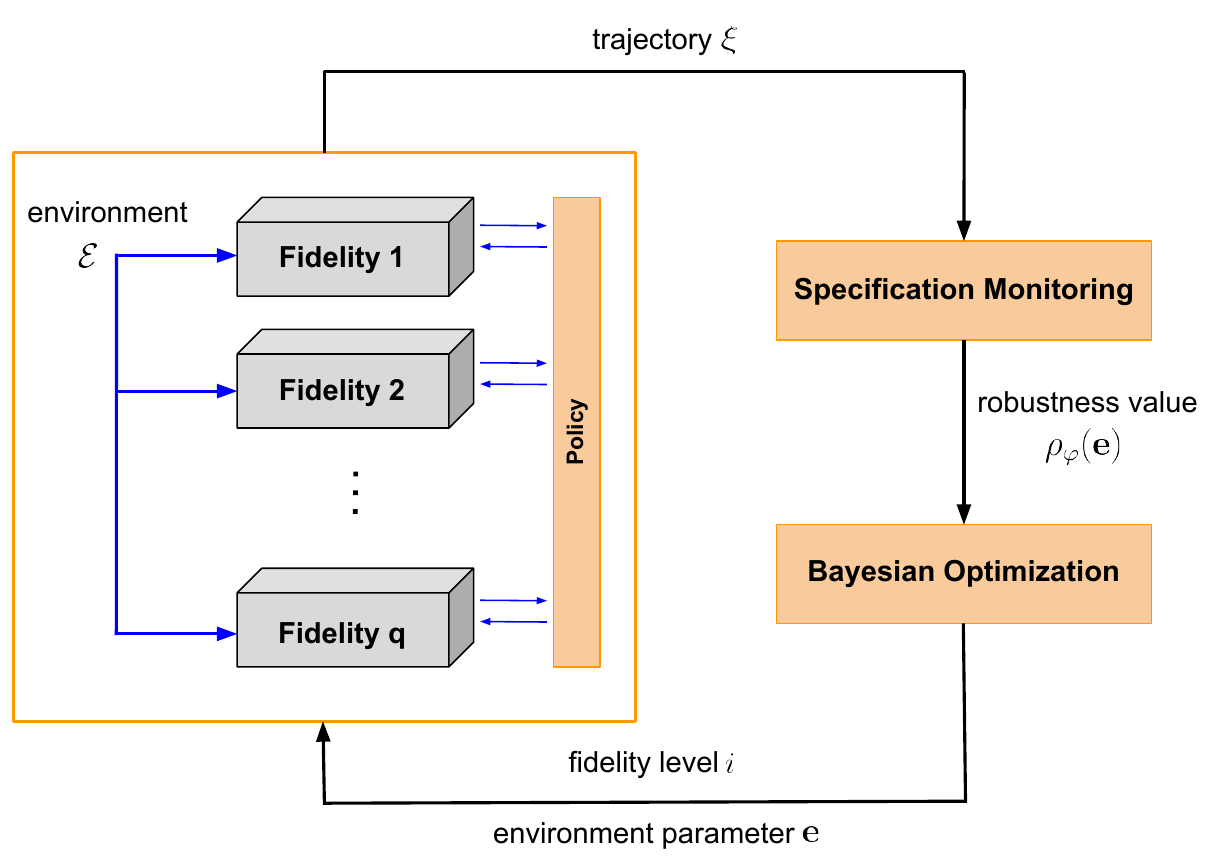}
     \caption{Overview of multi-fidelity BO falsification. The algorithm begins by performing random experiments at different fidelity levels, with fewer experiments conducted at higher fidelity levels. The robustness values across trajectories are recorded, and a GP model is initialized. The algorithm then optimizes entropy search over this model to select a candidate environment parameter $\mathbf{e}$, along with the fidelity index $i\in \left \{1,\ldots ,q  \right \}$ at which the next experiment should be performed. The GP model is then updated with the new data.}\label{Fig 1. }
\end{figure}

The quantitative semantics of the predicates are used to establish the quantitative semantics of the safety specification $\varphi$ using the following rules:
\begin{align}
\rho _{\mu > 0} &:= \mu \left ( \xi_{S} \left ( \mathbf{e} \right )  \right ) \label{(3)} \\
\rho _{\neg \mu}\left ( \xi_{S} \left ( \mathbf{e} \right )  \right ) &:= -\mu \left ( \xi_{S} \left ( \mathbf{e} \right )  \right ) \label{(4)}\\
\rho _{\varphi \wedge\psi  }\left ( \xi_{S} \left ( \mathbf{e} \right )  \right ) &:= \min \left ( \rho _{\varphi }\left ( \xi_{S} \left ( \mathbf{e} \right )  \right ), \rho _{\psi  }\left ( \xi_{S} \left ( \mathbf{e} \right )  \right )\right ) \label{(5)}\\
\rho _{\varphi \vee\psi  }\left ( \xi_{S} \left ( \mathbf{e} \right )  \right ) &:= \max \left ( \rho _{\varphi }\left ( \xi_{S} \left ( \mathbf{e} \right )  \right ), \rho _{\psi  }\left ( \xi_{S} \left ( \mathbf{e} \right )  \right )\right ) \label{(6)}
\end{align}

By evaluating the robustness value for a specific environment parameter $\rho _{\varphi }\left ( \xi_{S} \left ( \mathbf{e} \right ) \right )$, we can determine whether the system behaves safely in the corresponding $\mathbf{e}$. The sign of $\rho _{\varphi }\left ( \xi_{S} \left ( \mathbf{e} \right ) \right )$ serves as an indicator of whether the specification $\varphi$ is satisfied (non-negative value) or not satisfied (negative value). Additionally, robust semantics enables us to establish a hierarchy among failure cases. When $\rho _{\varphi} (\xi_{S}\left ( \mathbf{e}_{1} \right )) < \rho _{\varphi} (\xi_{S}\left ( \mathbf{e}_{2} \right ))$, failure case $\mathbf{e}_{1}$ is comparatively more severe than $\mathbf{e}_{2}$.
To simplify the notation, we use $\rho _{\varphi} (\mathbf{e})$ instead of $\rho _{\varphi }\left ( \xi_{S} \left ( \mathbf{e} \right ) \right )$ throughout the rest of paper.

\subsection{Bayesian Optimization}
There have been several applications of BO for testing learning-based control systems. For black-box systems in general, it is not known in advance how the robustness value $\rho _{\varphi} (\mathbf{e})$ depends on the environment parameter $\mathbf{e}$. To address this, we use GPs to estimate evaluations of robustness values within the environment parameters based on prior evaluations to aid in the optimization \cite{williams2006gaussian}. GPs are defined by a mean function $m: \mathbb{R}^n \to \mathbb{R}$ and a positive semi-definite covariance kernel function $k: \mathbb{R}^n \times \mathbb{R}^n \to \mathbb{R}$. In this manner, the unknown nonlinear robustness function $\rho _{\varphi} (\mathbf{e}):\mathcal{E}\rightarrow \mathbb{R}$ is represented as random variables, such that any finite number of them follows a multivariate Gaussian distribution. In this paper, the prior mean of the distribution $m(\mathbf{e})$ is set to $0$. The kernel function $k(\mathbf{e},\mathbf{e}^{\prime})$ expresses the covariance between function values $\rho _{\varphi }(\mathbf{e})$ and $\rho _{\varphi }(\mathbf{e}^{\prime})$ at two environment parameters $\mathbf{e}$ and $\mathbf{e}^{\prime}$. Suppose that we have a set of $n$ observations: $\mathbf{y}_{n}=[\hat{\rho}_{\varphi }(\mathbf{e}_{1}),\hat{\rho}_{\varphi }(\mathbf{e}_{2}),\ldots ,\hat{\rho}_{\varphi }(\mathbf{e}_{n})]$ observed with independent Gaussian noise $\omega \sim \mathcal{N}(0,\sigma ^{2})$ at environment parameters $\mathcal{E}_{n}=[\mathbf{e}_{1},\mathbf{e}_{2}, \ldots,\mathbf{e}_{n}]$, i.e., $\hat{\rho}_{\varphi }(\mathbf{e})=\rho_{\varphi }(\mathbf{e})+\omega$. By treating the outputs as random variables, the posterior distribution of $\rho _{\varphi }(\mathbf{e})$ is characterized by
\begin{align}
    m_{n}(\mathbf{e})&=\mathbf{k}_{n}(\mathbf{e})(\mathbf{K}_{n}+\mathbf{I}_{n}\sigma ^{2})^{-1}\mathbf{y}_{n} \label{(7)} \\
    k_{n}(\mathbf{e},\mathbf{e}^{\prime})&=k(\mathbf{e},\mathbf{e}^{\prime})-\mathbf{k}_{n}(\mathbf{e})(\mathbf{K}_{n}+\mathbf{I}_{n}\sigma ^{2})^{-1}\mathbf{k}_{n}^{T}(\mathbf{e}^{\prime}) \label{(8)}\\
    \sigma _{n}^{2}(\mathbf{e})&=k_{n}(\mathbf{e},\mathbf{e}^{\prime}) \label{(9)}
\end{align}
noting that the vector $\mathbf{k}_{n}(\mathbf{e}) =\left [ k(\mathbf{e},\mathbf{e}_{1}),\ldots ,k(\mathbf{e},\mathbf{e}_{n}) \right ]$, $\sigma _{n}^{2}(\mathbf{e})$ is variance, $\mathbf{I}_{n}$ is the identity matrix, and $\mathbf{K}_{n}$ is the positive definite kernel matrix $[k(\mathbf{e},\mathbf{e}^{\prime})]_{\mathbf{e},\mathbf{e}^{\prime}\in \mathcal{E}_{n} }$.

Search for an optimum in a GP is guided by an acquisition method, such as probability of improvement \cite{kushner1964new}, 
expected improvement \cite{movckus1975bayesian}, upper confidence bound \cite{srinivas2010gaussian}, and entropy search \cite{hennig2012entropy}. These functions balance between exploring new areas and exploiting promising ones. In areas where the objective function is predicted to be optimal or in areas that have not yet been explored, the acquisition function assigns a higher value.
\begin{algorithm*}[ht]
\caption{Multi-Fidelity BO for falsification}\label{alg:cap}
\begin{algorithmic}[1]

\Require simulators $S_{i}$, relative costs of simulators $\lambda _{i}$, $i=1,\ldots,q $
\Statex uncertainty space $\mathcal{E}$, policy $\pi$, safety specification $\varphi$, number of BO iterations $n$
\State Initialize GP over the nested training inputs from $q$ simulators and corresponding robustness values
\For{$t=1,2,\ldots,n $}
  \State Compute the acquisition function $\alpha^{\textup{ES}}(\mathbf{e})$ over GP model, as defined by Eq. \ref{(10)} \vspace{4pt}
  \State $\mathbf{e}^{\ast},i=\underset{\mathbf{e}\in \mathcal{E},i\in \left \{1, \ldots, q \right \}}{\mathrm{argmax}}\, \alpha^{\textup{ES}}_{i}\left ( \mathbf{e} \right )/\lambda _{i}$\vspace{4pt}
  \State Execute the scenario on $S_{i}$ for input $\mathbf{e}^{\ast}$ and monitor the resulting trajectory $\xi _{S_{i}}$
  \State Derive the safety specification robustness value $\rho_{\varphi}^{i}(\mathbf{e}^{\ast})$ across the trajectory
  \State Update GP model with new data
  \If{$\rho_{\varphi}^{i}(\mathbf{e}^{\ast})<0$}
         \State Counterexample $\mathbf{e}^{\ast}$ detected on fidelity $i$
  \EndIf
\EndFor
\end{algorithmic}
\end{algorithm*}

Entropy search, as an information-theoretic acquisition function, seeks to maximize the expected information gain from observing a subsequent query regarding the optimum, denoted as $\mathbf{e}^{\ast}$. We represent the $n$th probability distribution over $\mathbf{e}^{\ast}$ as $P_{n}(\mathbf{e}^{\ast})$ and its associated level of uncertainty as its entropy, denoted by $H(P_{n}(\mathbf{e}^{\ast}))$. Similarly, we can define $H(P_{n}(\mathbf{e}^{\ast}\mid \bar{\mathbf{e}},\bar{\rho}_{\varphi }\left ( \bar{\mathbf{e}} \right )))$ as the entropy of what the $\left ( n+1 \right )$th probability distribution over $\mathbf{e}^{\ast}$ would be if we were to observe it at an environment instance represented by $\bar{\mathbf{e}}$ and obtain a corresponding value of $\bar{\rho}_{\varphi }\left ( \bar{\mathbf{e}} \right )$. This measure of uncertainty depends on the specific observation value $\bar{\rho}_{\varphi }\left ( \bar{\mathbf{e}} \right )$. We can quantify the reduction in uncertainty resulting from making such an observation using
\begin{equation}
\alpha^{\textup{ES}}(\bar{\mathbf{e}})=H(P_{n}(\mathbf{e}^{\ast}))-\mathbb{E}\left[ H\Bigl(P_{n}\bigl(\mathbf{e}^{\ast}\mid\bar{\rho}_{\varphi }\left ( \bar{\mathbf{e}} \right )\bigl)\Bigl) \right ] \label{(10)}
\end{equation}
where $\mathbb{E}$ represents the expected value. Maximizing $\alpha^{\textup{ES}}(\bar{\mathbf{e}})$ translates to minimizing the uncertainty in our posterior after incorporating new observations.

\section{Methodology}
\label{sec:MFBO}

Full-fidelity simulators for a system of interest offer a variety of settings, the adjustment of which can significantly impact the simulator's fidelity. Starting with a high-fidelity simulator, lower-fidelity versions can be developed using techniques such as model order reduction, physical simplification, or data-driven approximations. Model order reduction simplifies the system's equations while preserving key dynamics, physical simplification approximates or ignores certain phenomena to expedite simulations, and data-driven approaches employ surrogate models to replicate high-fidelity outputs at reduced computational costs. In this work, we rely on domain expertise to develop low-fidelity simulators customized specifically to our application. Therefore, we consider $q$ simulators with different levels of fidelity for a system of interest specified by $S_{1}, \ldots,S_{q}$. Here, $S_{1}$ represents the lowest fidelity simulator (the cheapest, fastest, and least accurate one), while $S_{q}$ is the highest fidelity simulator (the most expensive, slowest, and most accurate one).

The relative query costs associated with these simulators, expressed by $\lambda _{1}, \ldots, \lambda _{q}$, where $1 < \lambda _{1}<\ldots < \lambda _{q}$ are determined by conducting some specific measurements. For each simulator, we can monitor trajectories and derive robustness values $\rho_{\varphi}^{i}\left (  \mathbf{e} \right )$ for $i\in \left \{1, \ldots, q \right \}$. Our goal is to estimate the black-box function $\rho_{\varphi}^{q}\left ( \mathbf{e} \right )$ given the observation information across different fidelities. Our first step is to establish a relationship between the robustness values obtained from various simulators. For this purpose, we use multi-output GPs that allow us to model a set of vector-valued outputs (the robustness values) based on a corresponding set of inputs (environment parameters) and represent the outputs as a multivariate normal distribution \cite{kennedy2000predicting}. For using multi-output GPs, we also add a dimension to the input space to represent the fidelity index $i\in \left \{1,\ldots ,q  \right \}$. Doing so, the multi-output GP can be seen as a single-output GP on the extended input space. Let $E_{i}$ be the environment configurations of fidelity $i$ for initializing GP model. To improve computational efficiency \cite{le2014recursive}, we suppose that these initialization input data have a nested structure, i.e., the initialization input data for a simulator with a higher fidelity index is a subset of the input data for a simulator with a lower fidelity index ($E_{q}\subseteq E_{q-1}\subseteq \ldots \subseteq E_{1}$). We assume that there is a linear correlation between the robustness values of different fidelity levels, as it facilitates straightforward integration of data from multiple fidelities. We use the auto-regression model \cite{kennedy2000predicting} to express the relation between fidelities
\begin{equation}
\rho_{\varphi}^{i}(\mathbf{e})=\eta_{i}  \rho_{\varphi}^{i-1}(\mathbf{e})+\rho _{gap}^{i}(\mathbf{e})
\label{(11)}
\end{equation}
Here, $\eta_{i}$ is a constant parameter that transfers knowledge linearly from the lower fidelity level to the higher one, and $\rho _{gap}^{i}(\mathbf{e})$ is the bias term between fidelities, which is an independent GP with its own mean function $m _{gap}^{i}$ and kernel function $k _{gap}^{i}(\mathbf{e},\mathbf{e}^{\prime})$. We assume $\rho_{\varphi}^{i-1}(\mathbf{e})$ and $\rho _{gap}^{i}(\mathbf{e})$ are independent processes linked only by the above equation. In this equation, while the linear term captures the overall trend between fidelity levels, the GP gap term is capable of capturing any residual non-linear relationships. These robustness values take the joint Gaussian distribution of the form
\begin{equation}
\begin{bmatrix}
\rho _{\varphi }^{i-1}
\\ \rho _{\varphi }^{i}
\end{bmatrix}\sim GP\Big(\begin{bmatrix}\textbf{0}
\\ \textbf{0}

\end{bmatrix},\begin{bmatrix}k_{i-1}
 & \eta_{i}  k_{i-1} \\ \eta_{i}  k_{i-1}
 & \eta_{i} ^{2}k_{i-1}+k_{gap}^{i}
\end{bmatrix}\Big) \label{(12)}
\end{equation}
The choice of kernel function is problem-dependent and encodes assumptions about smoothness and the rate of change of the robustness values. We use the radial basis function (RBF) kernel for both the error $k_{gap}^{i}$ and lower fidelity simulator indicated by $k_{i-1}$.

It is worth noting that considering linear relations between the outputs of different fidelity levels can simplify the computational modeling process but it is often insufficient for complex models. There are versions of multi-fidelity BO that assume nonlinear relationships between fidelities, as mentioned in \cite{perdikaris2017nonlinear}. However, these approaches also have limitations for use in certain contexts. One significant limitation is that these methods require lower-fidelity levels even during prediction to maintain a nested training set, which can be restrictive. Additionally, they often assume that the outputs are noiseless, which is challenging to meet.

Upon constructing the GP model, we use an acquisition function to select both the fidelity index and the environment parameter that leads to the minimum robustness value. We use a modified entropy search acquisition function:
\begin{equation}
\mathbf{e}^{\ast},i=\underset{\mathbf{e} \in \mathcal{E},i\in \left \{1,\ldots ,q  \right \}}{\mathrm{argmax}}\, \alpha^{\textup{ES}}_{i}(\mathbf{e})/\lambda _{i}\label{(13)}
\end{equation}
Eq. \ref{(13)} allows us to switch between different fidelity levels to reduce the computational costs of our experiments. Additionally, it performs two exploration and exploitation trade-offs at the same time: one at each fidelity level to search over that specific fidelity, and the other between different fidelity levels to build a more accurate GP model.

\begin{table*}[t]
\centering
\caption{Multi-fidelity settings for case studies}
\label{tab:fidelities}
\begin{tabular}{@{}m{2.5cm}m{3.5cm}m{2cm}rrr@{}}
\toprule
\textbf{Case Study} & \textbf{Setting} & \textbf{Type} & \textbf{Low-fidelity} & \textbf{Middle-fidelity} & \textbf{High-fidelity}\\
\midrule
\multirow{5}{*}{Cart-Pole} & Kinematic integrator & Categorical & Euler & Euler & semi-implicit Euler \\
 & Force magnitude & Continuous & $10$ &  $15$ & $20$\\
 & Position sensor Noise & Continuous & $\mathcal{N}(0, 0.25)$ & N/A & N/A\\
 & Sensors precision & Discrete &  $2$ digits &  $6$ digits &  $8$ digits\\
 & Episode length & Discrete & $150$ & $300$ & $450$\\
\midrule
\multirow{3}{*}{Lunar Lander} & Wind power & Continuous & $0$ & $6$ & $19.9$ \\
 & Turbulence power & Continuous & $0$ & $0$ &  $1.99$\\
 & Episode length & Discrete & $200$ & $400$ & $800$\\
\midrule
\multirow{4}{*}{Highway} & Simulation frequency & Discrete & $11$ & $13$ & $15$ \\
 & Number of road cars & Discrete & $23$ & $24$ & $25$\\
 & Position sensor noise & Continuous & $\mathcal{N}(0.1, 0.2)$ & $\mathcal{N}(0.1, 0.1)$ & N/A\\
 & Speed sensor noise & Continuous & $\mathcal{N}(0.5, 0.5)$ & $\mathcal{N}(0.3, 0.3)$ & N/A\\
\midrule
\multirow{3}{*}{Merge} & Max comfort acceleration & Continuous & $3$ & $3.5$ & $4$\\
 & Max comfort deceleration & Continuous & $5$ & $5.25$ & $5.5$\\
 & Time gap & Continuous & $1.5$ & $1.25$ & $1$\\
\midrule
\multirow{3}{*}{Roundabout} & Max imposed braking & Continuous & $1.8$ & $3$ & $4$\\
 & Lane change delay & Continuous & $0.84$ & $0.82$ & $0.8$\\
 & Politeness & Continuous & $0.1$ & $0.05$ & $0$\\
\bottomrule
\end{tabular}
\end{table*}

Algorithm \ref{alg:cap} outlines multi-fidelity BO for falsification. The algorithm is designed to model the robustness values using high-accuracy data from simulators with higher fidelity levels, as well as the data richness of simulators with lower fidelity levels. Once the prior GP over these $q$ simulators is obtained, BO is performed iteratively until a stopping criterion is met. On each iteration, the algorithm seeks to find the worst counterexample by maximizing the acquisition function in Eq. \ref{(13)}. Depending on the gap and cost ratios between simulators, the algorithm chooses the candidate environment parameter and fidelity index to run the next experiment. Then, the GP is updated with the new environment instance and its corresponding robustness value on that specific simulator. The algorithm outputs the identified counterexamples and their corresponding fidelity indices. Fig. \ref{Fig 1. } shows multi-fidelity BO approach applied to falsification.

\section{Experiments}
\label{sec:Evaluation}
In this section, we provide the implementation details and outline the methodology used to measure the cost ratios between different fidelity levels. We then discuss the baselines used for comparison. Finally, we describe our benchmarks. For each of these case studies, we identify the variables that influence the fidelity of simulator. These variables can be of various types: discrete, continuous, or categorical. Once these variables are identified, we explore different combinations of these variables to adjust and control the simulator's fidelity. We develop low-, middle-, and high-fidelity simulators and examine both two- and three-fidelity BO approaches. The source code for our experiments is  online.\footnote{\href{https://github.com/SAILRIT/MFBO_Falsification}{\textcolor{blue}{https://github.com/SAILRIT/MFBO\_Falsification}}}


\subsection{Cost Ratios Measurement}
\label{sec:COST}

Selecting the next fidelity level for evaluation in Eq. \ref{(13)} depends on the relative costs of simulators. Hence, we avoid allocating pre-determined constant costs to simulators and consider that the cost ratios of the high-fidelity simulator to the other two simulators can be determined by measuring two factors: simulation time and output similarity. Simulation time measures how long it takes for simulators to run a similar experiment, and output similarity quantifies how closely the trajectories of different simulators align. Numerous tests are available to determine the similarity between trajectories of the low-, middle-, and high-fidelity simulators subjected to identical inputs. However, we employ cosine similarity due to its straightforwardness and clarity. Eq. \ref{(14)} demonstrates the cost ratios between high-, low-, and middle-fidelity simulators:
\begin{equation}
    \begin{aligned}
        \frac{\lambda_{hf}}{\lambda_{lf}} &= \frac{t_{hf}}{t_{lf}} \times s_{lf}^{hf} \\
        \frac{\lambda_{hf}}{\lambda_{mf}} &= \frac{t_{hf}}{t_{mf}} \times s_{mf}^{hf}
    \end{aligned}
    \label{(14)}
\end{equation}
where $t_{hf}$, $t_{mf}$, and $t_{lf}$ are the time for executing an identical experiment on the high-fidelity, low-fidelity, and middle-fidelity simulators. Further, cosine similarities between trajectories of the high-fidelity simulator and trajectories of the low- and middle-fidelity simulators are
\begin{equation}
    \begin{aligned}
        s_{lf}^{hf} &= \frac{\xi_{hf} \cdot \xi_{lf}}{\|\xi_{hf}\| \|\xi_{lf}\|} \\
        s_{mf}^{hf} &= \frac{\xi_{hf} \cdot \xi_{mf}}{\|\xi_{hf}\| \|\xi_{mf}\|}
    \end{aligned}
    \label{(15)}
\end{equation}

To provide more precise cost ratios, we compute the average similarity among ten trajectories for each of our case studies, along with the mean simulation time resulting from ten identical experiments conducted at various fidelity levels.
\begin{figure}[!t]
  \centering
  \begin{subfigure}{0.235\textwidth}
    \includegraphics[width=\linewidth]{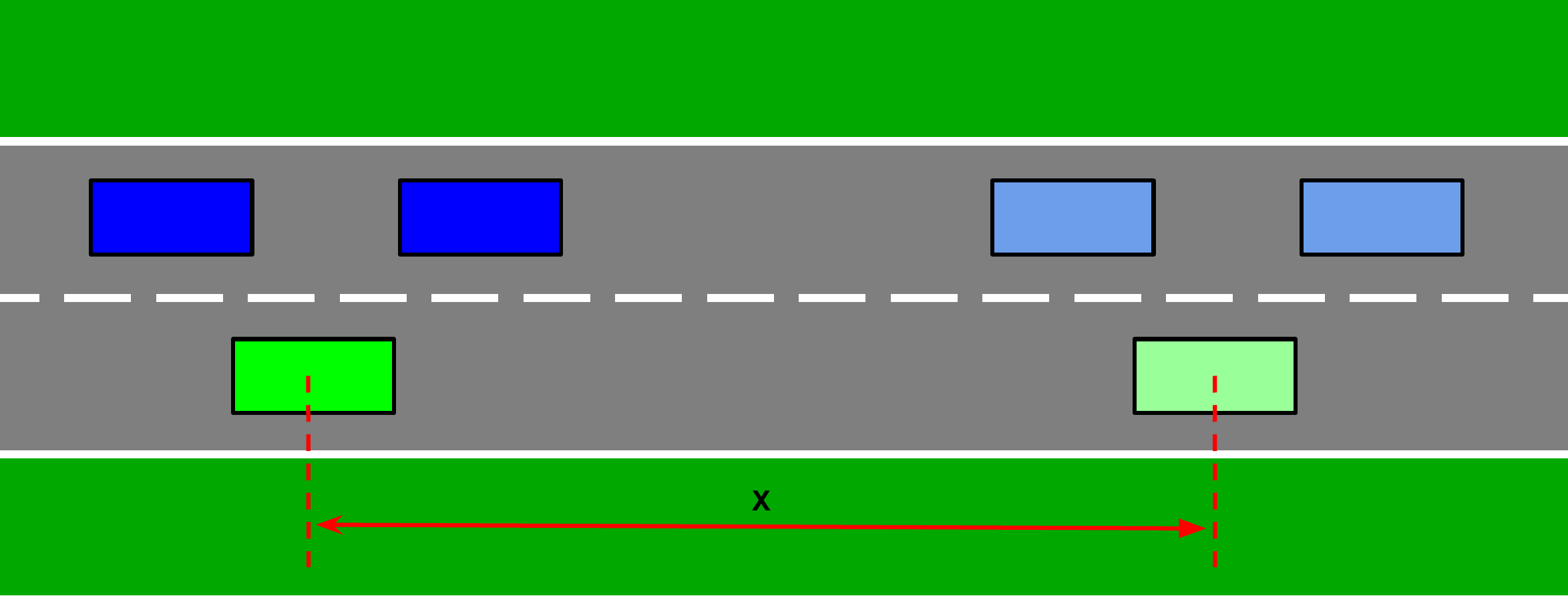}
    \caption{ }
    \label{fig:subfig-a}
  \end{subfigure}
  \hfill
  \begin{subfigure}{0.235\textwidth}
    \includegraphics[width=\linewidth]{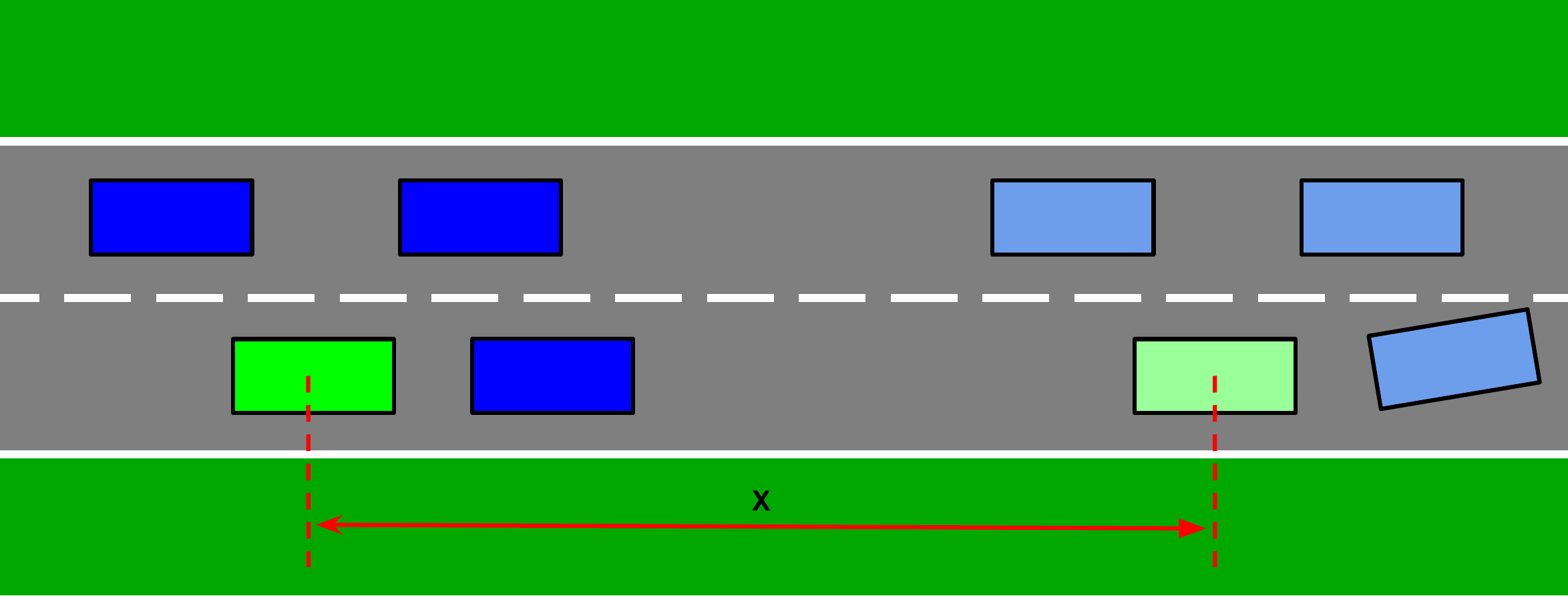}
    \caption{}
    \label{fig:subfig-b}
  \end{subfigure}
  \caption{Fidelity levels in the highway case study. (a) Low-fidelity highway simulator at $t=t_{0}$ and $t=12t_{0}$. (b) High-fidelity highway simulator at $t=t_{0}$ and $t>16t_{0}$. The simulation frequencies in the low-fidelity simulator and the high-fidelity simulator are $11$ and $15$; respectively. For an identical scenario, the high-fidelity simulator is slower. Additionally, the high-fidelity simulator features more cars on the road.}
  \label{Fig 2. }
\end{figure}

\subsection{Baselines}
We compare the performance of multi-fidelity BO with standard BO across high-, middle-, and low-fidelity simulators. In addition to these individual fidelity levels, we included two other baselines in our analysis: TuRBO-$1$ and $\pi$BO.

\subsubsection{\textbf{TuRBO-\texorpdfstring{$1$}{1}}}

TuRBO-$1$\footnote{\href{https://github.com/uber-research/TuRBO/blob/master/turbo/turbo_1.py}{https://github.com/uber-research/TuRBO}} is a variant of the TuRBO algorithm \cite{eriksson2019scalable} that focuses on using a single trust region for optimization. In TuRBO-$1$, the optimization is performed within a hyperrectangular trust region, which is centered at the current best solution, denoted by $\mathbf{e}^{\ast}$. To balance between exploration and exploitation, the side lengths of this region are adjusted based on the success of previous optimization steps. In fact, the trust region's size $L$ is initially set to $L_{\text{init}}$. It expands or contracts according to the number of consecutive successful or unsuccessful evaluations. Specifically, after $\tau_{\text{succ}}$ successful evaluations, the region's size is doubled, $L \leftarrow \min(L_{\text{max}}, 2L)$. Conversely, after $\tau_{\text{fail}}$ unsuccessful evaluations, the size is reduced by half, $L \leftarrow L/2$. If the size $L$ falls below a minimum threshold $L_{\text{min}}$, the trust region is terminated and reinitialized. The acquisition function $\alpha(\mathbf{e}, D)$, where $D$ denotes the dataset of observations, guides the selection of new points within the trust region, and the optimization continues until convergence or a predefined budget is reached. The flexibility of TuRBO-$1$ to change the trust region size helps avoid getting stuck in local optima and increases optimization process efficiency. In our implementation, the trust region size is initialized to $L_{\text{init}} = 0.75$, can expand to a maximum of $L_{\text{max}} = 1.5$, and is reduced down to a minimum of $L_{\text{min}} = 0.5^7$ when necessary.
\begin{figure*}[ht!]
    \centering
     \includegraphics[scale=0.46]{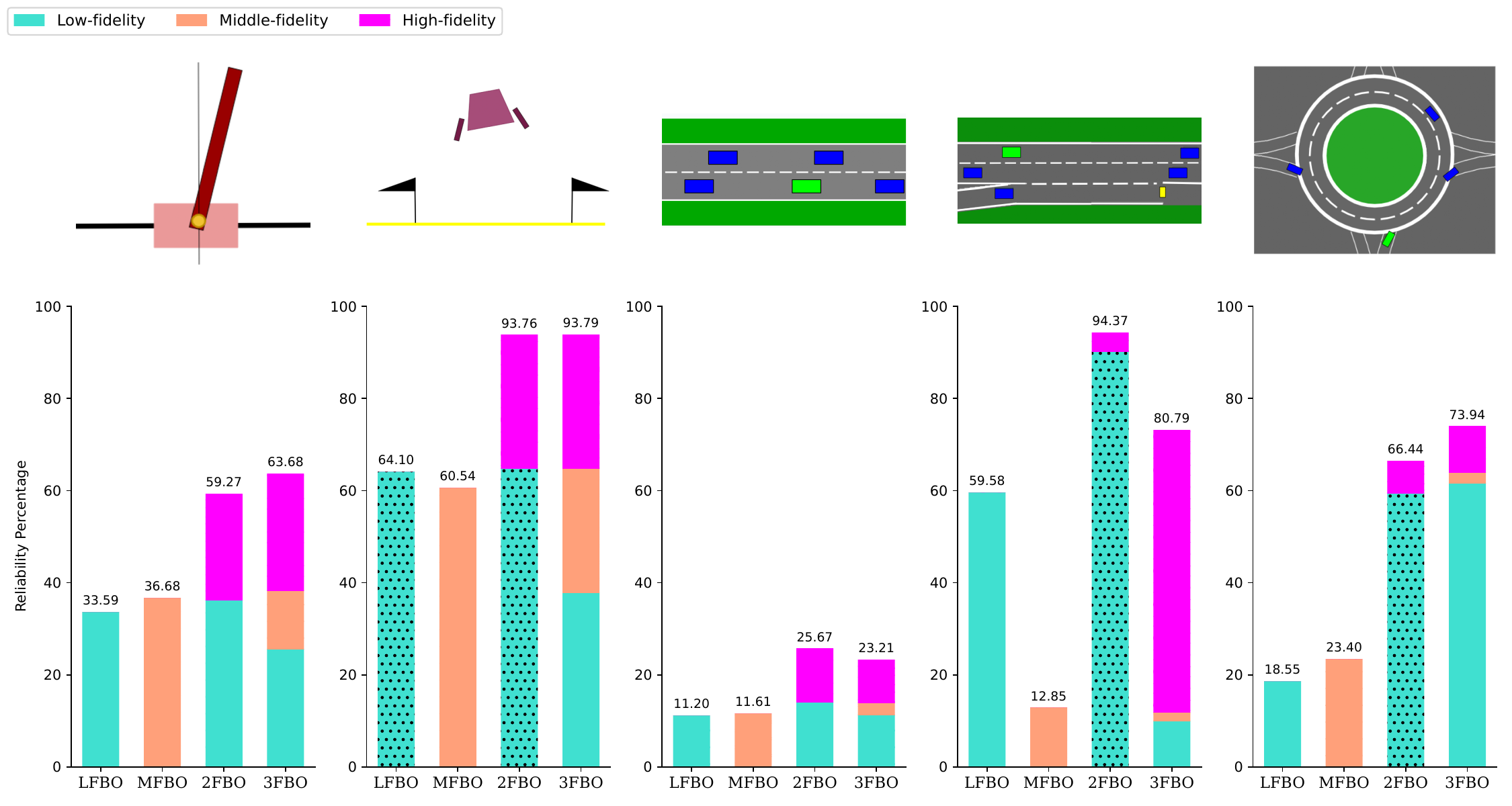}
     \caption{Average reliability percentage of discovered counterexamples over $750$ tests using standard BO on low- and middle-fidelity simulators, two-fidelity BO and three-fidelity BO methods for cart-pole, lunar lander, highway, merge, and roundabout case studies.}
     \label{Fig 3. }
\end{figure*}

\subsubsection{\textbf{\texorpdfstring{$\pi$BO}{piBO}}}

In $\pi$BO \cite{hvarfner2022pi}, expert knowledge is represented as a probability distribution $\pi(\mathbf{e})$ over the potential location of the optimum. This prior information is used to weight the acquisition function and guide the optimization process towards promising regions. The acquisition function $\alpha(\mathbf{e}, D_{n-1})$, where $D_{n-1} = \{(\mathbf{e}_1, \rho_{\varphi}(\mathbf{e}_1)), \ldots, (\mathbf{e}_{n-1}, \rho_{\varphi}(\mathbf{e}_{n-1}))\}$, is modified by multiplying it with $\pi(\mathbf{e})^{\beta^*/n}$, where $\beta^*$ is a hyperparameter indicating the expert's confidence in the prior knowledge, and $n$ represents the current BO iteration. This is mathematically represented as:
\begin{equation}
\mathbf{e}^{\ast} \in \arg\max_{\mathbf{e} \in \mathcal{E}} \alpha(\mathbf{e}, D_{n-1}) \cdot \pi(\mathbf{e})^{\beta^*/n}
\end{equation}
As more data is collected through iterations, the exponent $\beta^*/n$ becomes smaller, which reduces the impact of the prior $\pi(\mathbf{e})$ and shifts the focus to the data-driven surrogate model. For $\pi$BO baseline, we consider the falsified points to be located on the edges of uncertainty space \cite{ramezani2021testing}. Therefore, the prior knowledge of expert is a U-shaped distribution where the edges are weighted more than the inside area. In practical terms, we modeled the prior distribution at each dimension of uncertainty space as a combination of two Gaussian distributions, each centered at the edges. The value of $\beta^*$ is set to $0.7$.

The rationale for selecting TuRBO-$1$ and $\pi$BO as baselines in our study is based on their common foundation in BO and their unique strategies for improving optimization performance. TuRBO-$1$ serves as a baseline to investigate whether we should concentrate on high-fidelity models to enhance optimization or incorporate information from uncertain sources to potentially improve efficiency and reduce computational costs. On the other hand, by contrasting our multi-fidelity method with $\pi$BO, we aim to highlight the benefits of using available lower-fidelity models instead of relying solely on expert knowledge, especially in situations where such knowledge may be limited or uncertain.

\subsection{Case Studies}
To evaluate our proposed approach, we implemented multi-fidelity BO using Emukit toolkit \cite{emukit2023} and conducted experiments on five benchmarks provided by Gym \cite{brockman2016openai}. For each case study, we elaborate on safety specifications, system controller, and how different levels of fidelity are developed. For policy training, we used OpenAI baselines
\footnote{\href{https://github.com/openai/baselines}{\textcolor{blue}{https://github.com/openai/baselines}}} on the high-fidelity simulator.

\subsubsection{\textbf{Classic Control---Cart-Pole}}

The cart-pole environment comprises a cart and a vertical pole attached to the cart using a passive pivot joint. The objective of this task is to keep the vertical pole balanced while the cart moves to the right or left along a frictionless track. The state vector of the system consists of the cart position $x$, cart velocity $v$, pole angle $\theta $, and pole angular velocity $\dot{\theta }$. The agent can perform two actions: apply force to move the cart left or right. The environment comes with six sources of uncertainty: $\mathcal{E}\ = \left [-2,2 \right ]\times \left [-0.05, 0.05  \right ] \times \left [ -0.2, 0.2 \right ]\times \left [ -0.05, 0.05 \right ]\times \left [0.05, 0.15  \right ]\times \left [0.4, 0.6  \right ]$. The first four uncertainty intervals represent the possible perturbations to the position, velocity, angle, and angular velocity of the system. Following that, the last two intervals are for the mass and length of the pole, respectively. A system's trajectory is a series of states across time, i.e., $\xi=(x(t),v(t),\theta(t),\dot{\theta }(t))$. Given an instance $\mathbf{e}\in \mathcal{E}$, the trajectory of the system is uniquely defined.

We consider the primary cart-pole offered by Gym as low-fidelity simulator, which has a force magnitude of $10$. To create a high-fidelity simulator, we attempt to make the game more difficult and unstable by increasing the force magnitude applied to the pole. We also use a stronger strategy to solve the kinematic equations using the semi-implicit Euler method rather than Euler method for the low-fidelity simulator. The parameters chosen for multi-fidelity experiments are shown in Table~\ref{tab:fidelities}. Moreover, as per our measurements, the cost ratios amount to $\lambda_{hf} \slash \lambda_{mf}= 2.71$ and $\lambda_{hf} \slash \lambda_{lf}= 20.81$. We trained a controller for the high-fidelity simulator using proximal policy optimization (PPO) \cite{schulman2017proximal}.

The controller for this case study must meet certain safety requirements. First, the cart position $x$ should be within the range $\left [ -1,1 \right ]$ ($\mu _{1}$). Second, the absolute momentum of the cart should be less than $1$ ($\mu _{2}$). Third, the cart-pole angle should be less than $9$ degrees from vertical ($ \mu _{3}$). Hence, the specification could be written as: $\varphi = \mu _{1}\wedge \mu _{2} \wedge \mu _{3}$.

\subsubsection{\textbf{Classic Control---Lunar Lander}}

This environment features a lander that must safely descend onto the moon's surface under conditions of low gravity. The primary purpose is to guide the agent to the landing pad as softly and fuel-efficiently as possible. For this environment, there are eight state variables including the coordinates $\left ( x,y \right )$ of the lander, the horizontal and vertical velocities $\left( v_{x},v_{y}\right )$, the orientation $\theta$ in space, the angular velocity $v_{\theta}$, and two Boolean parameters indicating whether the left or right leg has made contact with the ground. Four possible actions include activating the left orientation engine, activating the right orientation engine, activating the main engine, and taking no action. We consider four sources of uncertainty. The coordinate perturbations are such that $\delta _{x}\in \left [ -0.5,0.5 \right ]$ and $\delta _{y}\in \left [ 0,3 \right ]$, the velocity perturbations are such that $\delta _{v_{x}}\in \left [-2,2 \right]$ and $\delta _{v_{y}}\in \left [0,2 \right]$. A trajectory of this system could be represented as $\xi=(x(t),y(t),v_{x}(t),v_{y}(t),\theta(t),\dot{\theta }(t))$.

The high-fidelity simulator uses two sources of wind disturbance within the environment. In the middle-fidelity simulator, we replicate one of these disturbances, while the low-fidelity counterpart disregards both disturbances. The low-fidelity simulator includes inaccurate sensors capable of providing measurements up to two decimal points. The detailed distinctions in fidelity levels for the lunar lander case study are presented in Table~\ref{tab:fidelities}. Based on this configuration, the cost ratios are $\lambda_{hf} \slash \lambda_{mf} = 2.02$ and $\lambda_{hf} \slash \lambda_{lf} = 4.53$.

We employ the deep deterministic policy gradient (DDPG) algorithm \cite{lillicrap2015continuous} to train a controller for the high-fidelity simulator. A trajectory is safe when it satisfies these conditions: First, the lander must keep its horizontal coordinate ($x$) within a certain range of the origin ($\mu_{1}$). Second, the lander's tilt angle must not exceed $\pi /4$ ($\mu_{2}$), and third, the lander does not spin faster than $0.2$ radians per second ($\mu_{3}$). Thus, the safety specification is $\varphi =\mu _{1} \wedge \mu _{2}\wedge \mu _{3} $.

\begin{figure}[ht!]
    \centering
     \includegraphics[scale=0.55]{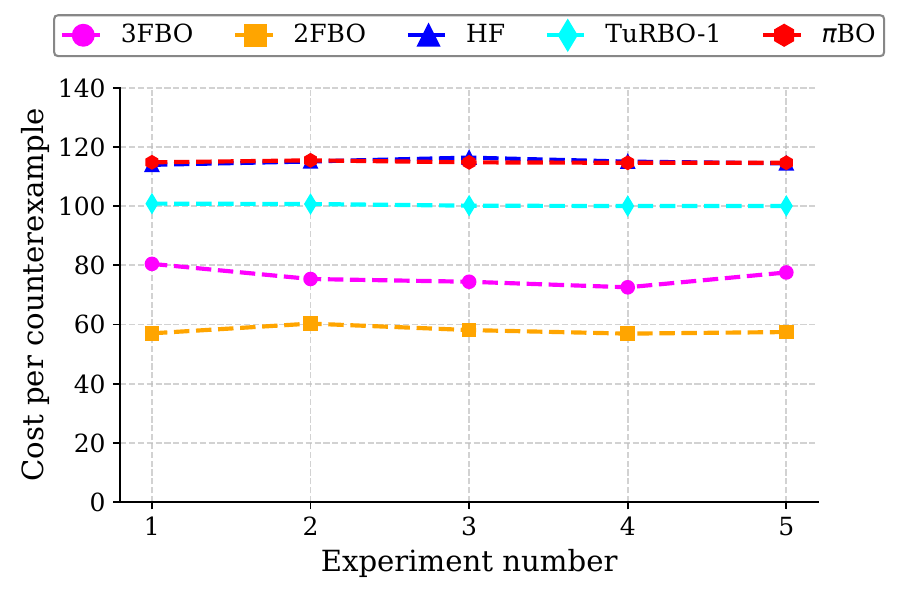}
     \caption{ Comparison between average cost of finding a counterexample through $200$ BO iterations on cart-pole case study.}
     \label{Fig 4. }
\end{figure}

\subsubsection{\textbf{Autonomous Driving---Highway, Merge, and Roundabout Scenarios}}

To thoroughly evaluate the effectiveness of our approach, we concentrate on three driving scenarios: highway driving, merging, and roundabout navigation. In these scenarios, the state space includes the $\left ( x,y \right )$ longitudinal and lateral coordinates, as well as longitudinal and lateral velocities $\left( v_{x},v_{y}\right )$ of both the ego vehicle and the four closest cars. We consider five discrete meta-actions: moving to the left lane, maintaining an idle state, shifting to the right lane, accelerating, and decelerating. The trajectories of these scenarios comprise the coordination of both the ego vehicle and the four nearest cars, represented as
\begin{multline*}
\xi=(x_{ego}(t),y_{ego}(t),x_{ob_{1}}(t),y_{ob_{1}}(t), \\
x_{ob_{2}}(t),y_{ob_{2}}(t),x_{ob_{3}}(t),y_{ob_{3}}(t),x_{ob_{4}}(t),y_{ob_{4}}(t)).
\end{multline*}

For these scenarios, we use deep Q-learning (DQN) \cite{mnih2015human} for the ego vehicle and both intelligent driver model (IDM) \cite{treiber2000congested} and MOBIL (minimizing overall braking induced by lane change) model \cite{kesting2007general} for longitudinal and lateral behaviors of other vehicles on the road, respectively. While MOBIL is a lane-changing model that uses safety and incentive criteria—considering the braking of the new and old neighbors, as well as a politeness factor—to determine when a lane change is safe and beneficial, IDM is a parametric car-following model that regulates the acceleration of vehicles. In this model, the vehicle's acceleration is described by the following kinematic equation:
\begin{equation}
\dot{v}_{\alpha} = a \left[ 1 - \left( \frac{v_{\alpha}}{v_{0}} \right)^{4} - \left( \frac{s^{*}(v_{\alpha}, \Delta v_{\alpha})}{s_{\alpha}} \right)^{2} \right]
\end{equation}
where the acceleration $\dot{v}_{\alpha}$ depends on the current speed $v_{\alpha}$, the gap to the leading vehicle $s_{\alpha}$, and the relative speed difference $\Delta v_{\alpha}$ between the ego vehicle and the vehicle ahead. The deceleration factor is influenced by the ratio of the desired minimum gap $s^{*}(v_{\alpha}, \Delta v_{\alpha})$ to the actual gap $s_{\alpha}$, where $s^{*}(v_{\alpha}, \Delta v_{\alpha}) = s_{0} + vT + \frac{v \Delta v}{2\sqrt{ab}}$. Here, $s_0$ represents the minimum gap in dense traffic, $vT$ is the gap maintained while following the leading vehicle at a constant time gap $T$, and $a$ and $b$ correspond to the maximum comfortable acceleration and deceleration, respectively.

\noindent{\textbf{Highway:}} This environment involves a vehicle on a multi-lane highway \cite{highway-env}. The agent controls speed and lane transitions. The aim is to achieve high speeds without colliding with adjacent vehicles while maintaining a position on the right side of the road. The motion of the ego vehicle is represented using kinematic bicycle model \cite{polack2017kinematic}.  
We evaluate nine potential sources of uncertainty, arising from the initial proximity of the ego vehicle to its four nearest neighbors as well as their initial velocities. The initial spacing between the ego vehicle and the four nearest vehicles falls within the range $\delta_x \in [20, 30]$. Meanwhile, the ego vehicle's initial velocity is limited to the range $v_{ego} \in [20, 40]$. Similarly, the initial velocities of other four cars are denoted as $v_{ob_{i}}\in [20, 30]$ for $i= 1, \ldots, 4$.

\begin{figure}[ht!]
    \centering
     \includegraphics[scale=0.55]{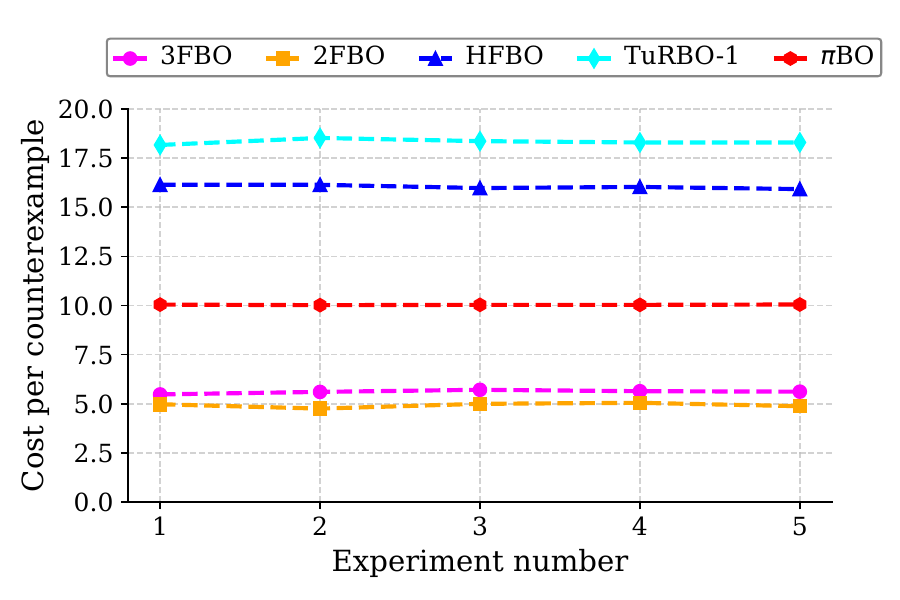}
     \caption{ Comparison between average cost of finding a counterexample through $200$ BO iterations on lunar lander case study.}
     \label{Fig 5. }
\end{figure}

We developed our simulators to have different levels of fidelity. The high-fidelity simulator, with a simulation frequency of $15$, is more accurate than the low- and middle-fidelity simulators, which operate at frequencies of $11$ and $13$, respectively. Our low- and middle-fidelity simulators introduce noise into the sensors, limiting the ego vehicle's ability to precisely predict the location and velocity of neighboring vehicles while also maintaining fewer cars on the road, as shown in Fig. \ref{Fig 2. }. Based on these settings, we found the cost ratios to be $\lambda_{hf} \slash \lambda_{mf}= 3.75$ and $\lambda_{hf} \slash \lambda_{lf}= 14.5$. Safe trajectories are established by ensuring that an absolute distance of over $0.5$ meters is maintained between the ego car and both the leading (front) and following (back) vehicles positioned in the same line. Accidents involving other vehicles on the road, not including the ego vehicle, are not considered unsafe modes. Hence, our safety specification can be written as $\varphi = \left ( \left | x_{ego}-x_{ob_{i}} \right |-0.5 \right )$ for $i= 1, \ldots, 4$. 

\begin{figure}[ht!]
    \centering
     \includegraphics[scale=0.55]{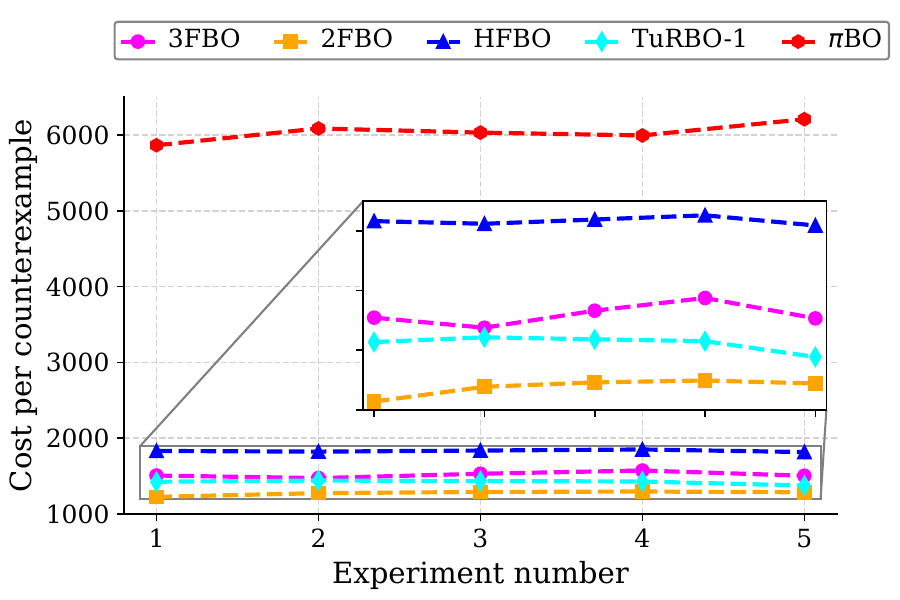}
     \caption{ Comparison between average cost of finding a counterexample through $200$ BO iterations on highway driving environment.}
     \label{Fig 6. }
\end{figure}

\noindent{\textbf{Merge:}} This scenario simulates a highway merge negotiation task where the ego-vehicle approaches a merge point with vehicles incoming from an access ramp. The agent is rewarded for maintaining high speed, avoiding collisions, and making space for merging vehicles. The agent's goal is to maintain a high speed while safely accommodating other vehicles during the merge. We evaluate nine potential sources of uncertainty, focusing on the initial velocities of the ego vehicle and the four closest cars, within the range $\left[ 8, 12 \right]$, as well as perturbations on the initial positions of these four nearby vehicles in range $\left[ 10, 12 \right]$.
To develop simulators with varying levels of fidelity for this scenario, we adjust the IDM parameters for the vehicles on the road, including maximum comfortable acceleration and deceleration, to create a more aggressive driving environment. The most aggressive scenario is designated as the high-fidelity simulator, and the controller is trained specifically for this environment. For the low- and middle-fidelity simulators, we modify the parameters to create more conservative scenarios that also include sensor noise. The specific design choices for the fidelity settings are detailed in Table~\ref{tab:fidelities}. Based on these settings, we obtained the cost ratios to be $\lambda_{hf} \slash \lambda_{mf}= 1.2$ and $\lambda_{hf} \slash \lambda_{lf}= 2.6$. A trajectory is considered safe if the ego vehicle maintains a minimum distance of $5$ meters from the nearest car in the same lane throughout the trajectory.

\noindent{\textbf{Roundabout:}}
In this task, the ego-vehicle is navigating a roundabout with flowing traffic. While it follows a pre-determined route automatically, it must manage lane changes and control its speed to pass the roundabout quickly while avoiding collisions. We evaluate five potential sources of uncertainty, arising from the vehicle length $l\in [5, 7]$ and perturbations in the initial positions of the four nearest neighbors $\delta_x \in [-5, 5]$.
We developed three levels of fidelity by modifying key parameters of the MOBIL model. The low-fidelity level used parameters that promote more conservative driving, such as a higher politeness factor and increased lane-change delays. In contrast, the high-fidelity level employed parameters with quicker lane-change decisions and stricter acceleration and braking constraints. Based on these configurations, we determined the cost ratios to be $\lambda_{hf} \slash \lambda_{mf}= 4.6$ and $\lambda_{hf} \slash \lambda_{lf}= 27.2$. The safety requirements for this case study dictate that the differences in both the x-coordinates and y-coordinates between the ego vehicle and other vehicles should not simultaneously fall below certain thresholds. Specifically, the y-coordinate difference should not fall below 4.4 units ($\mu_1$) at the same time that the x-coordinate difference drops below 2.1 units ($\mu_2$).

\section{Results and Discussion}
\label{sec:result}

In this section, we evaluate various aspects of multi-fidelity BO on the case studies we discussed previously. To examine the established trade-off between accuracy and cost by multi-fidelity BO framework, we begin by assessing the accuracy through the reliability of counterexamples suggested by multi-fidelity BO methods and standard BO on the low- and middle-fidelity simulators. We then discuss the cost-effectiveness of multi-fidelity BO in comparison to BO on high-fidelity, TurBO-$1$, and $\pi$BO baselines.

\begin{figure}[ht!]
    \centering
     \includegraphics[scale=0.55]{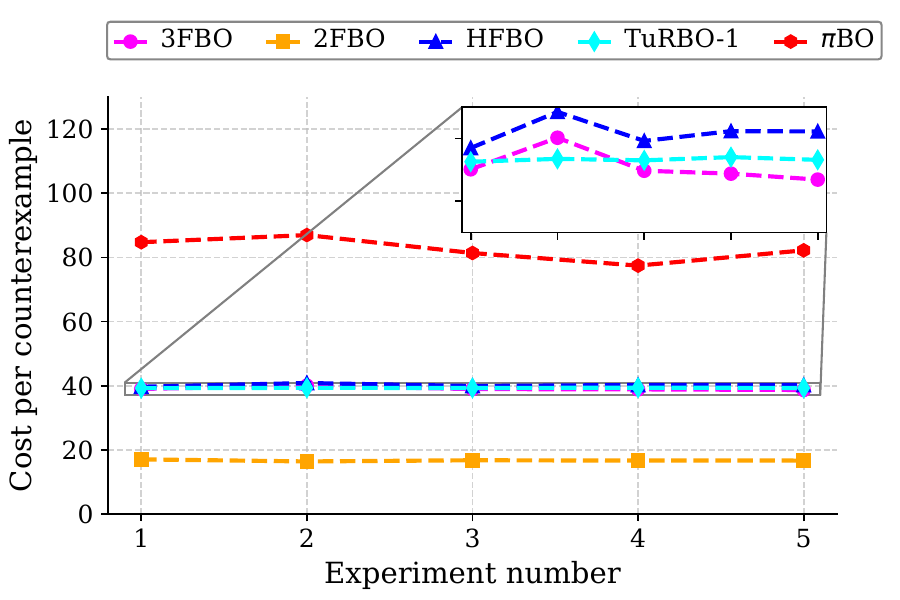}
     \caption{Comparison between average cost of finding a counterexample through $200$ BO iterations on merge case study.}
     \label{Fig 7. }
\end{figure}


We run $5$ experiments, each consisting of the average results from $150$ tests over $200$ BO iterations. We validate the counterexamples provided by lower-fidelity simulators in multi-fidelity BO methods and the counterexamples identified by conducting BO on the low- and middle-fidelity simulators. We define the reliability percent as the ratio of real counterexamples to the total number of identified counterexamples. As depicted in Fig. \ref{Fig 3. }, the average reliability percentages for both two- and three-fidelity BO methods on the cart-pole case study are about $26\%$ higher than BO on the low- and middle-fidelity simulators. This improvement for the lunar lander case study is about $30\%$ while for the highway driving environment it is about $12\%$. More significant results hold in the case of the merge and roundabout scenarios. To this end, using multi-fidelity BO yields more accurate counterexamples compared to using solely lower-fidelity simulators.

Comparing two-fidelity and three-fidelity BO methods, the impact of the middle-fidelity simulator on accuracy becomes evident. In the cart-pole case study, transitioning from a two-fidelity setting to a three-fidelity setting resulted in a $27$ reduction in the use of the low-fidelity simulator, with a corresponding increase in the use of the middle-fidelity simulator, as detailed in Table~\ref{tab:executions}. Since this simulator provides more reliable counterexamples in comparison to the low-fidelity simulator, three-fidelity BO demonstrates increased reliability compared to two-fidelity BO. The roundabout scenario yields similar but more remarkable results. In the lunar lander case study, despite conducting 
$30\%$ of the simulations on the middle-fidelity simulator, the transition from a two-fidelity to a three-fidelity setting does not enhance reliability. This lack of improvement is due to the minimal difference between the middle-fidelity and low-fidelity simulators, as also reflected in their similar relative costs. In the highway driving case study, the three-fidelity BO method does not provide more accurate counterexamples. This can have several reasons: First, $2\%$ of accurate simulations on high-fidelity were replaced by inaccurate ones from middle-fidelity which results in decreased reliability of three-fidelity BO compared to two-fidelity BO. Second, the accuracy of the multi-fidelity model becomes questionable as the dimension of the uncertainty space increases, similar to the behavior observed in the merge scenario, which further reduces the effectiveness of the three-fidelity approach.
\begin{table}[ht]
    \centering
    \begin{tabular}{@{}lrrrr@{}}
        \toprule
        Case study & $2$F/$3$F & High-fidelity & Middle-fidelity & Low-fidelity \\
        \midrule
        \multirow{2}{*}{Cart-pole} & $2$F & $23.3$ & N/A & $76.7$  \\
         & $3$F & $23$ & $27.4$ & $49.6$  \\
        \midrule
        \multirow{2}{*}{Lunar Lander} & $2$F & $27.2$ & N/A & $72.8$  \\
         & $3$F & $29.3$ & $30.6$ &  $40.1$ \\
        \midrule
        \multirow{2}{*}{Highway} & $2$F & $10.6$ & N/A & $89.4$  \\
         & $3$F & $8.6$ & $18$ & $73.4$  \\
        \midrule
        \multirow{2}{*}{Merge} & $2$F & $8.3$ & N/A & $91.7$  \\
         & $3$F & $50$ & $31.1$ & $18.9$  \\
        \midrule
        \multirow{2}{*}{Roundabout} & $2$F & $12.6$ & N/A & $87.4$  \\
         & $3$F & $14$ & $8.9$ & $77.1$  \\
        \bottomrule
    \end{tabular}
    \caption{Average percentage of executions on high-fidelity, middle-fidelity, and low-fidelity simulators in two- and three-fidelity BO settings over $750$ tests through $200$ BO iterations.}
    \label{tab:executions}
\end{table}

\begin{figure}[ht!]
    \centering
     \includegraphics[scale=0.55]{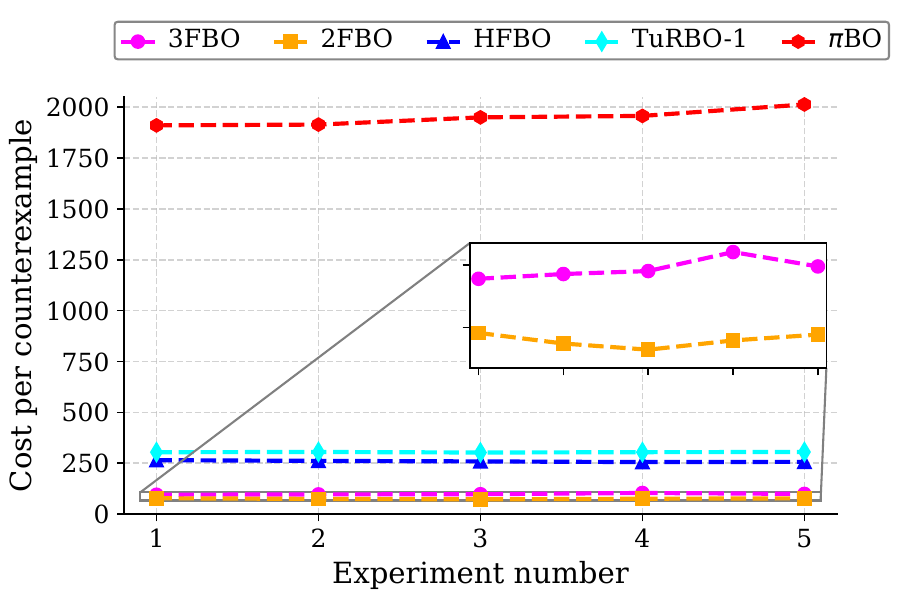}
     \caption{Comparison between average cost of discovering a single counterexample through $200$ BO iterations on roundabout scenario.}
     \label{Fig 8. }
\end{figure}

\begin{figure*}[ht!]
    \centering
     \includegraphics[scale=0.41]{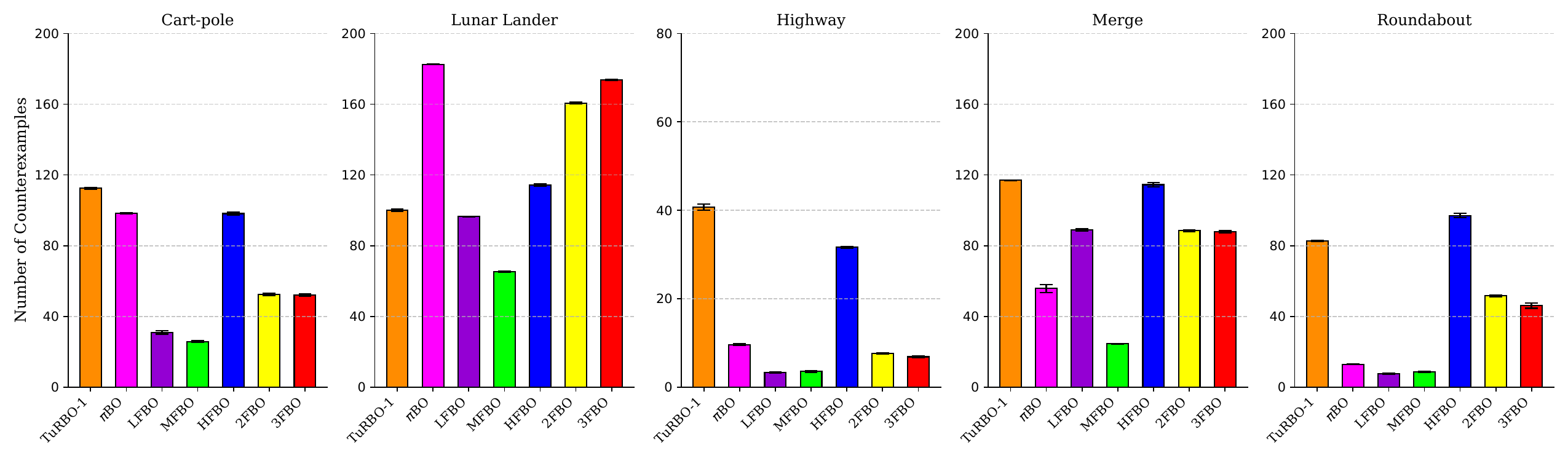}
     \caption{Comparison between number of counterexamples discovered by two- and three-fidelity BO methods, standard BO on single fidelities, TuRBO-$1$, and $\pi$BO baselines.}
     \label{Fig 9. }
\end{figure*}

In our previous work, we used multi-fidelity BO to solve the falsification problem across various case studies. The results indicated that multi-fidelity BO identified more counterexamples at the same cost compared to high-fidelity BO. Specifically, the results revealed that for the same cost, two-fidelity BO detected counterexamples $3.31$ and $1.5$ times more frequently than high-fidelity BO on the cart-pole and lunar lander, respectively. Our primary criterion was ensuring the algorithm's ability to transition between fidelity levels with pre-determined costs.

The performance of multi-fidelity BO is heavily influenced by the difference between simulators, which affects the cost ratios. This is especially important when selecting the next fidelity level to use. Therefore, we derived these cost ratios based on precise measurements on the simulators. Based on the validated counterexamples, we illustrated the average costs of identifying a single counterexample for standard BO, TuRBO-$1$, and $\pi$BO on the high-fidelity simulator in addition to two-fidelity and three-fidelity BO methods through $200$ BO iterations over $5$ experiments for the cart-pole, lunar lander, and autonomous driving case studies in Fig. \ref{Fig 4. }--\ref{Fig 8. }. As we expected, multi-fidelity BO consistently outperforms standard BO conducted on high-fidelity in all five case studies. In the cart-pole case study, using three- and two-fidelity BO could reduce the cost of falsification up to $49.5\%$ and $32.4\%$ in comparison to BO on the high-fidelity simulator. In the lunar lander environment, we observe $69\%$ and $65\%$ cost reduction, while these percentages for the highway case study are $18\%$ and $26\%$ by three- and two-fidelity BO methods, respectively. In the merge case study, the cost reduction percentages achieved with two- and three-fidelity BO are $2.7\%$ and $58\%$. Additionally, three- and two-fidelity BO methods in the roundabout case study offer $61\%$ and $70\%$ cost reduction. Hence, given the same cost, multi-fidelity BO can find more counterexamples compared to the high-fidelity BO. We also can see that the integration of the middle-fidelity and the shift from the two-fidelity setting to three-fidelity can increase the cost per counterexample because some evaluations on the low-fidelity simulator are replaced by more costly experiments on the middle-fidelity simulator, as shown in Table~\ref{tab:executions}. Our analysis showed the role of the middle-fidelity in three-fidelity BO cost reduction. In the three-fidelity setting, if the middle-fidelity simulator is very similar to the low-fidelity simulator and does not provide much new information, the cost of finding a single counterexample is close to the cost of finding a single counterexample with two-fidelity BO, as evident in the lunar lander case study. However, as the middle-fidelity simulator becomes more informative and accurate, the average cost of the three-fidelity BO method approaches that of full-fidelity BO as observed in cart-pole and merge case studies. In the merge case study, the algorithm tends to use higher-fidelity simulators more frequently, often disregarding the exploration of low-fidelity simulator, which leads to a greater cost difference compared to the two-fidelity setting.

Additionally, in comparison to TuRBO-$1$ and $\pi$BO, multi-fidelity BO finds lower cost per counterexample in all case studies except for the highway driving case, where TuRBO-$1$ outperforms three-fidelity BO. This is mainly because TuRBO-$1$ excels in higher dimensional search spaces, such as in the highway and merge scenarios. In contrast, in the three-fidelity BO for the highway driving case study, the fewest experiments were conducted on the high-fidelity simulator (unlike the merge scenario), which hindered the performance of the three-fidelity BO.

We also depicted number of real detected counterexamples by different methods in Fig. \ref{Fig 9. }. While in the cart-pole, highway, and merge case studies, TuRBO-$1$ outperforms other methods, in the lunar lander case study $\pi$BO provides more counterexamples. Moreover, in all case studies except for lunar lander, high-fidelity BO suggests more counterexamples with respect to multi-fidelity BO methods. In the lunar lander case study, because the gaps between low- and middle-fidelity simulators are not significant from the high-fidelity simulator, the counterexamples suggested by multi-fidelity BO methods are highly reliable, as we mentioned earlier. This results in multi-fidelity methods discovering more counterexamples than standard BO on the high-fidelity simulator for the same number of BO iterations. Conversely, in the cart-pole case study, where the middle-fidelity simulator can provide more reliable counterexamples compared to the low-fidelity simulator, three-fidelity BO detects more counterexamples than the two-fidelity BO method. Moreover, in the highway driving case study, where some high-fidelity experiments were replaced by those on the middle-fidelity simulator, three-fidelity BO identifies fewer counterexamples compared to the two-fidelity BO framework.

\noindent{\textbf{Limitations:}}
The proposed multi-fidelity BO approach, while promising, does have several limitations. A key challenge lies in accurately estimating the relative computational costs of different simulators. While we did not use arbitrarily chosen costs, inaccuracies in these estimates can negatively impact the optimization process, potentially reducing the effectiveness of our multi-fidelity approach. Additionally, the assumption of a linear relationship between robustness values, which are not continuous functions, while simplifying the computational process, may not be sufficient to capture the complexities of models and influence the efficiency of the framework. Another challenge is the identification and utilization of appropriate lower-fidelity simulators. It can be difficult to ensure that these simulators are sufficiently accurate to guide the optimization process toward global optimum.

\section{Conclusions}\label{sec: Conclusion}
We presented an extended study of multi-fidelity falsification of closed-loop control systems in a simulated environment. Our approach stands out for its ability to considerably cut down on computational cost by intelligently switching between simulators with varying levels of fidelity. To make multi-fidelity BO more robust, we used specific measurements on the simulators to determine the cost proportions of them with different fidelity levels. We demonstrated the effectiveness of our framework by applying it to controllers designed for Gym environments.

One promising future direction is to develop an adaptive strategy for transitioning to high-fidelity simulators to efficiently manage computational resources. This approach would involve using lower-fidelity simulators until they no longer provide significant improvements or cost-efficient insights in the optimization process, and then the algorithm would move to high-fidelity simulations. Another area for future exploration is enhancing the multi-fidelity approach to effectively manage situations where the operational environment or input variables differ across various levels of fidelity. This would use space mapping techniques to correlate the outputs of simulators operating under different conditions, which improves the robustness and applicability of the framework in real-world scenarios. Another direction for future research is the development of a cost-aware multi-fidelity falsification process. The algorithm can determine the optimal cost structure of the simulators throughout the optimization, offering a mechanism that is self-adjusting, especially if initial cost estimates are inaccurate.

\section*{Acknowledgements}

This research was supported in part by the National Science Foundation (NSF) under Award No. 2132060 and the Federal Aviation Administration (FAA) under Contract No. 692M15-21-T-00022.

\renewcommand*{\bibfont}{\footnotesize}
\printbibliography

\end{document}